\documentclass[twocolumn]{aastex63}

\shorttitle{Binary Cepheid with a 59-day orbital period}
\shortauthors{Pilecki et al.}
\graphicspath{{figs/}{./}}

\begin{document}

\title{Discovery of a binary-origin classical Cepheid in a binary system with a 59-day orbital period
\footnote{Based on observations collected at the European Southern Observatory, Chile}
\footnote{This paper includes data gathered with the 6.5m Magellan Clay Telescope at Las Campanas Observatory, Chile.}
}

\correspondingauthor{Bogumił Pilecki}
\email{pilecki@camk.edu.pl}

\author[0000-0003-3861-8124]{Bogumi{\l} Pilecki}
\affiliation{Centrum Astronomiczne im. Miko{\l}aja Kopernika, PAN, Bartycka 18, 00-716 Warsaw, Poland}


\author{Ian B. Thompson}
\affiliation{Carnegie Observatories, 813 Santa Barbara Street, Pasadena, CA 91101-1292, USA}

\author{Felipe Espinoza-Arancibia}
\affiliation{Centrum Astronomiczne im. Miko{\l}aja Kopernika, PAN, Bartycka 18, 00-716 Warsaw, Poland}

\author{Richard I. Anderson}
\affiliation{Institute of Physics, Laboratory of Astrophysics, EPFL, Observatoire de Sauverny, 1290 Versoix, Switzerland}

\author{Wolfgang Gieren}
\affiliation{Universidad de Concepci{\'o}n, Departamento de Astronom{\'i}a, Casilla 160-C, Concepci{\'o}n, Chile}

\author{Weronika Narloch}
\affiliation{Universidad de Concepci{\'o}n, Departamento de Astronom{\'i}a, Casilla 160-C, Concepci{\'o}n, Chile}

\author{Javier Minniti}
\affiliation{Centrum Astronomiczne im. Miko{\l}aja Kopernika, PAN, Bartycka 18, 00-716 Warsaw, Poland}

\author{Grzegorz Pietrzy{\'n}ski}
\affiliation{Centrum Astronomiczne im. Miko{\l}aja Kopernika, PAN, Bartycka 18, 00-716 Warsaw, Poland}

\author{M{\'o}nica Taormina}
\affiliation{Centrum Astronomiczne im. Miko{\l}aja Kopernika, PAN, Bartycka 18, 00-716 Warsaw, Poland}

\author{Giuseppe Bono}
\affiliation{Dipartimento di Fisica Universit`a di Roma Tor Vergata, viadella Ricerca Scientifica 1, 00133 Rome, Italy}

\author{Gergely Hajdu}
\affiliation{Centrum Astronomiczne im. Miko{\l}aja Kopernika, PAN, Bartycka 18, 00-716 Warsaw, Poland}

\begin{abstract}

We report the discovery of a surprising binary configuration of the double-mode Cepheid OGLE-LMC-CEP-1347 pulsating in the first (P$_1$=0.690d) and second overtone (P$_2$=0.556d) modes. The orbital period (P$_{orb}$=59d) of the system is five times shorter than the shortest known to date (310d) for a binary Cepheid. The Cepheid itself is also the shortest-period one ever found in a binary system and the first double-mode Cepheid in a spectroscopically double-lined binary.
OGLE-LMC-CEP-1347 is most probably on its first crossing through the instability strip, as inferred from both its short period and fast period increase, consistent with evolutionary models, and from the short orbital period (not expected for binary Cepheids whose components have passed through the red giant phase). Our evolutionary analysis yielded a first-crossing Cepheid with a mass in a range of 2.9-3.4 M$_\odot$ (lower than any measured Cepheid mass), consistent with observations.
The companion is a stable star, at least two times fainter and less massive than the Cepheid (preliminary mass ratio $q$=0.55), while also redder and thus at the subgiant or more advanced evolutionary stage. To match these characteristics, the Cepheid has to be a product of binary interaction, most likely a merger of two less massive stars, which makes it the second known classical Cepheid of binary origin.
Moreover, further evolution of the components may lead to another binary interaction.

\end{abstract}

\keywords{stars: variables: Cepheids - binaries: spectroscopic - stars: late-type}


\section{Introduction} \label{sec:intro}

Classical Cepheids (hereafter also Cepheids) are crucial for various fields of astronomy including stellar oscillations and the evolution of intermediate and massive stars, and with enormous influence on modern cosmology.  Since the discovery of the relationship between their pulsation period and luminosity (the Leavitt Law, \citealt{1912HarCi.173....1L}), the Cepheids have been extensively used to measure distances in the Universe. They are radially pulsating evolved intermediate and high-mass giants and supergiants, mostly located in a well defined position on the helium-burning loop (called the {\em blue loop}). Theory predicts masses of Cepheids in the range 3-11 M$_\odot$ \citep{Cox_1980_ARAA_CEP_masses,Bono_1999_ApJS_cep_masses_range_high,Bono_2001_ApJ_cep_masses_range_low, Anderson_2016_Rotation}, indicating that unless a binary interaction has occurred in their past, all progenitors are B-type stars \citep{Pecaut_2013_alltype_MS_data,Evans_2020_Cep_companions}. As such, they are mostly members of binary and multiple systems \citep{Moe_2017_multip_review}.

For low-mass ($\lesssim 3.5$ M$_\odot$) stars the blue loops predicted by evolution theory are too short to reach the instability strip and explain the existence of short-period Cepheids \citep{Anderson_2016_Rotation, DeSomma_2021_MNRAS_per-age}. Such low-mass stars cross the instability strip only during the earlier rapid subgiant phase of evolution (in the Hertzsprung gap), but only a few percent of Cepheids are expected to be found in that stage \citep{DeSomma_2021_MNRAS_per-age}. Nevertheless, these first-crossing Cepheids are currently the only possible explanation for the existence of short-period Cepheids \citep[see, e.g.][]{Ripepi_2022_MNRAS_VMC_LMC_CEPs}. The difficulty in proving this scenario arises mainly from the low number of accurate dynamical mass measurements of Cepheids. Only seven have been measured thus far \citep{allcep_pilecki_2018,Gallenne_2018_V1334Cyg_Cep}, and all of them occur in a very narrow mass range around 4 M$_\odot$. The lowest mass, 3.61 M$_\odot$, was measured for the binary Cepheid OGLE-LMC-CEP-4506 \citep{cep9009apj2015,allcep_pilecki_2018} in the Large Magellanic Cloud (LMC).

Binary systems with Cepheid stars are expected to have orbital periods longer than 200 days. These long periods are required so that the binaries survive the evolution of the components on the red giant branch \citep{Neilson2015_cep_binsys_porb, Moe_2017_multip_review}. However, the first-crossing Cepheids can have shorter periods, as they have not yet passed through this stage. At the moment the shortest period measured for a genuine binary Cepheid is 310 days for OGLE-LMC-CEP-0227 \citep{cep227nature2010,cep227mnras2013}, while in general periods of about 1 year or longer are being found.

We have recently started a project to identify and characterize a large sample of Cepheids in spectroscopically double-lined binary (SB2) systems, suitable for dynamical mass measurements. The survey is currently focused on the LMC galaxy but is being extended to other galaxies as well. In the first step of the project we have confirmed that virtually all Cepheids that lie significantly above the period-luminosity relation have a red giant companion easily detectable in the spectra \citep{cepgiant1_2021}. The next step is to confirm the orbital motion of these Cepheids and to determine the orbital periods of the systems.

In this paper we present the discovery of a surprising binary configuration of the Cepheid \href{http://simbad.u-strasbg.fr/simbad/sim-id?Ident=OGLE+LMC-CEP-1347}{OGLE-LMC-CEP-1347} (henceforth also LMC-CEP-1347), one of the 41 SB2 Cepheid candidates presented in \citet{cepgiant1_2021}. In Section~\ref{sec:spectro} we present the spectroscopic data that we have acquired together with the preliminary orbital solution. In Section~\ref{sec:discussion} we discuss the origin and the current state of the Cepheid and draw a few direct conclusions.

\section{Spectroscopic observations}
\label{sec:spectro}

\begin{figure}
    \begin{center}
        \includegraphics[width=1.0\linewidth]{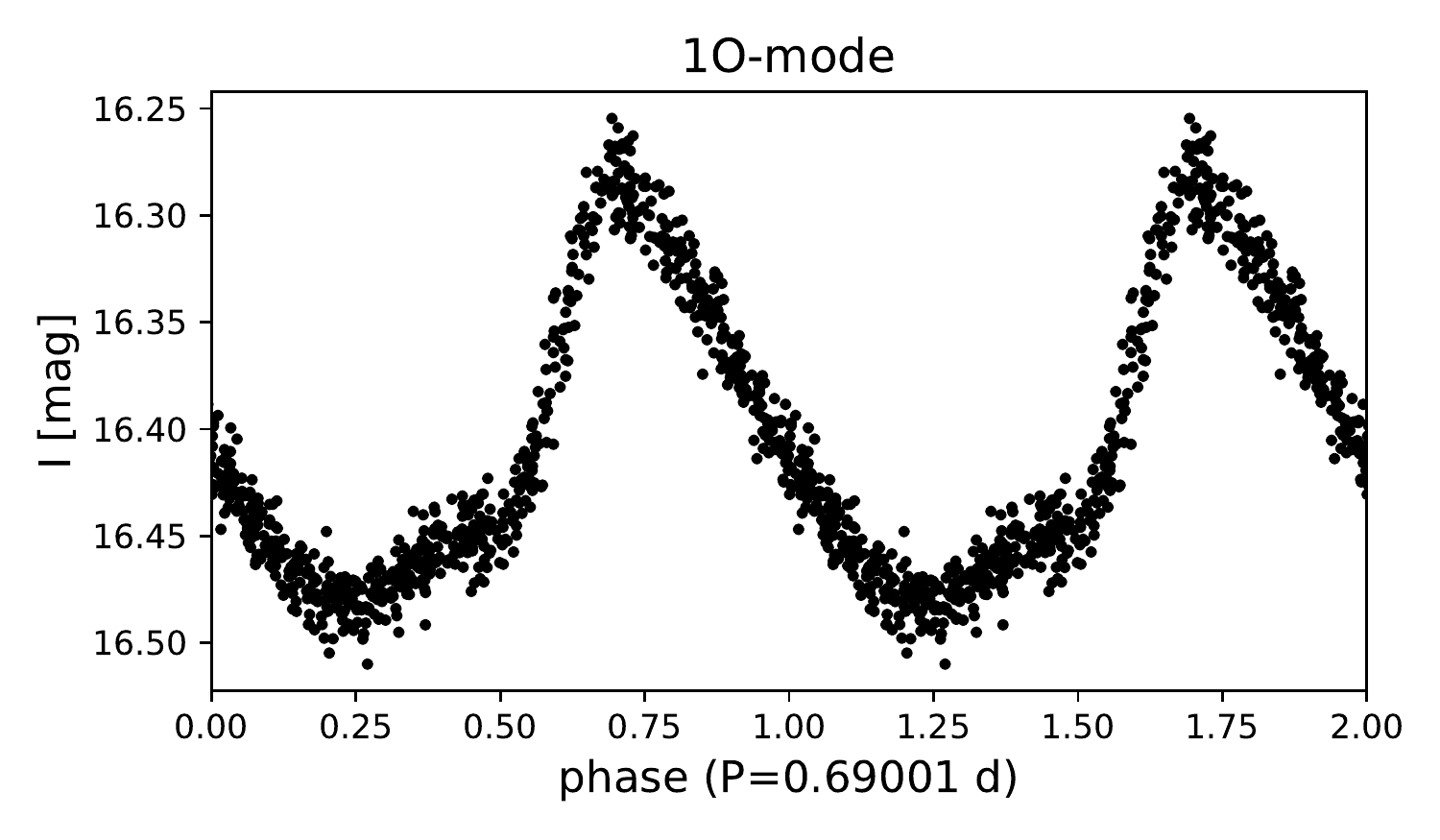}
        \includegraphics[width=1.0\linewidth]{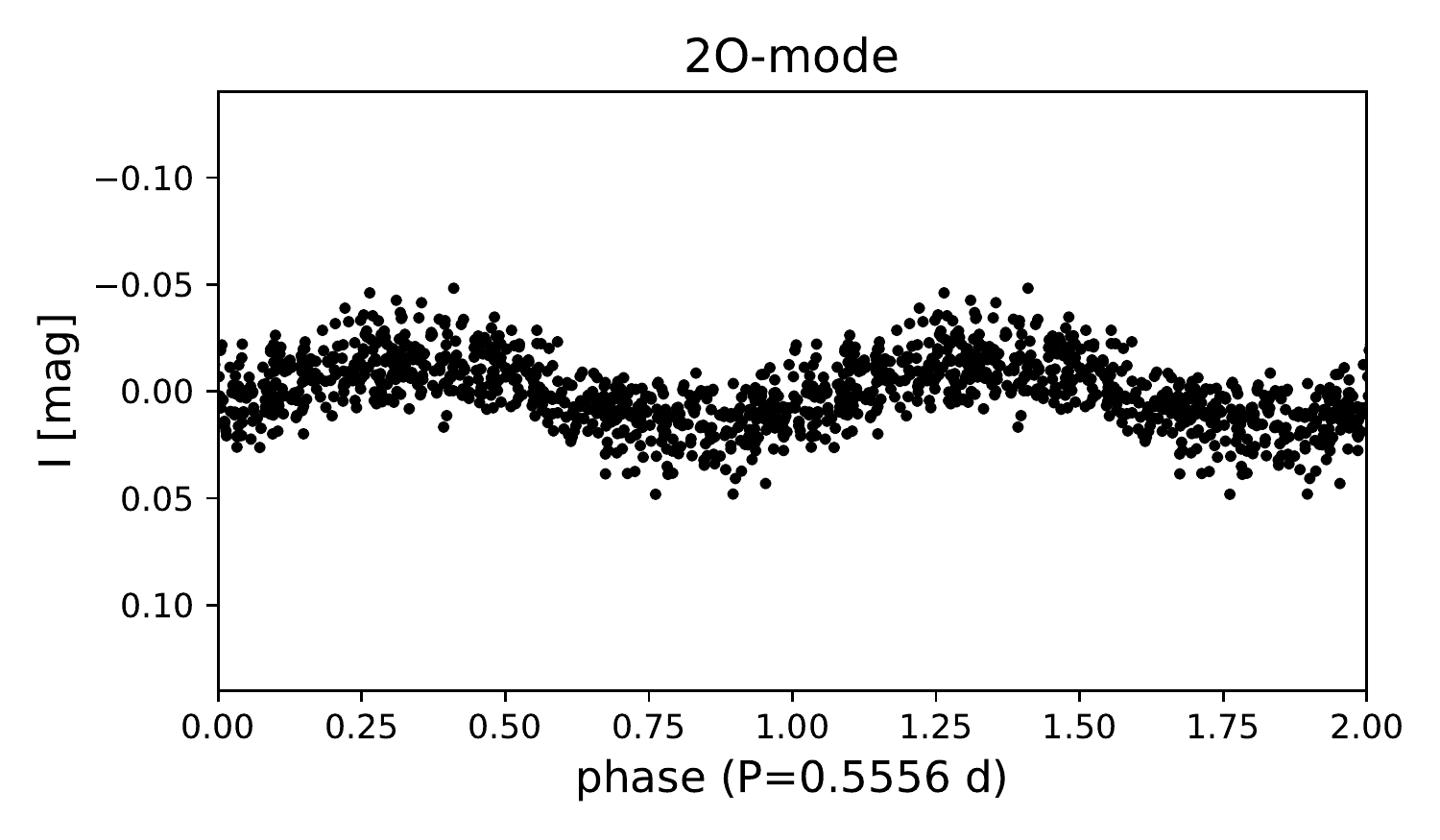}
    \end{center}
\caption{I-band light curves showing variability for the first ($1O$) and second overtone ($2O$) mode with the same Y-axis span. Data come from the OGLE project \citep{Soszynski_2017AcA_OCVS_MC_Cep} and were disentangled to show variability in each mode separately. }
\label{fig:lcs}
\end{figure}

LMC-CEP-1347 (I=16.4 mag, V=17.1 mag; $\alpha_{J2000}$=$05^{h}15^{m}06.4^{s}$, $\delta_{J2000}$=$-69^{\circ}39'53"$; \citealt{Soszynski_2017AcA_OCVS_MC_Cep}) is a double-mode Cepheid with first-overtone ($1O$) period of P$_1$=0.690 day and second-overtone (2O) period of P$_2$=0.556 day (P$_2$/P$_1$=0.805).
The I-band amplitude of this Cepheid is almost six times higher in the $1O$ mode than in the $2O$ mode -- see Fig.\ref{fig:lcs}.
We observed this target spectroscopically between October 2021 and January 2022 using two instruments, UVES at the 8.2m Very Large Telescope UT2 (Paranal, ESO) and MIKE at the 6.5m Clay Magellan telescope (Las Campanas, Carnegie) in Chile. Four spectra were taken with the UVES spectrograph (wavelength range 3300-6600\AA) in service mode. With an exposure time of 1400s, we obtained a typical signal-to-noise ratio S/N$\sim$10 at resolution R$\sim$60000. The spectra were reduced using the {\tt eso-reflex} software available at the ESO software repository\footnote{\url{https://www.eso.org/sci/software/pipelines/}}. Five spectra were taken with the MIKE spectrograph (3300-9600\AA) in remote mode with an exposure time of 1800s (typical S/N$\sim$15 at R$\sim$40000) and were reduced using Daniel Kelson's pipeline available at the Carnegie Observatories Software Repository\footnote{\url{http://code.obs.carnegiescience.edu}}.

We used the Broadening Function (BF) technique \citep{Rucinski_1992AJ_BF,Rucinski_1999ASPC_BF_SVD} implemented in the RaveSpan code \citep{t2cep098apj2017} to measure the radial velocities (RV) of the components of the binary.
This technique provides narrower velocity profiles than the cross-correlation function method, helping in the separation of components and increasing the accuracy of the RV measurements.
This is important as the companion is significantly fainter than the Cepheid. To measure the BF profiles we used a set of wide wavelength bands between 4100 \AA{} and 6600 \AA{}, omitting broad and atmospheric absorption lines. In 8 spectra the profiles of the Cepheid and its companion were clearly separated providing accurate measurements of the RVs of the individual components. For one MIKE spectrum the BF profiles overlap, but velocities of both components could still be measured (they differ by 16 km/s only).

\begin{figure}
    \begin{center}
        \includegraphics[width=1.0\linewidth]{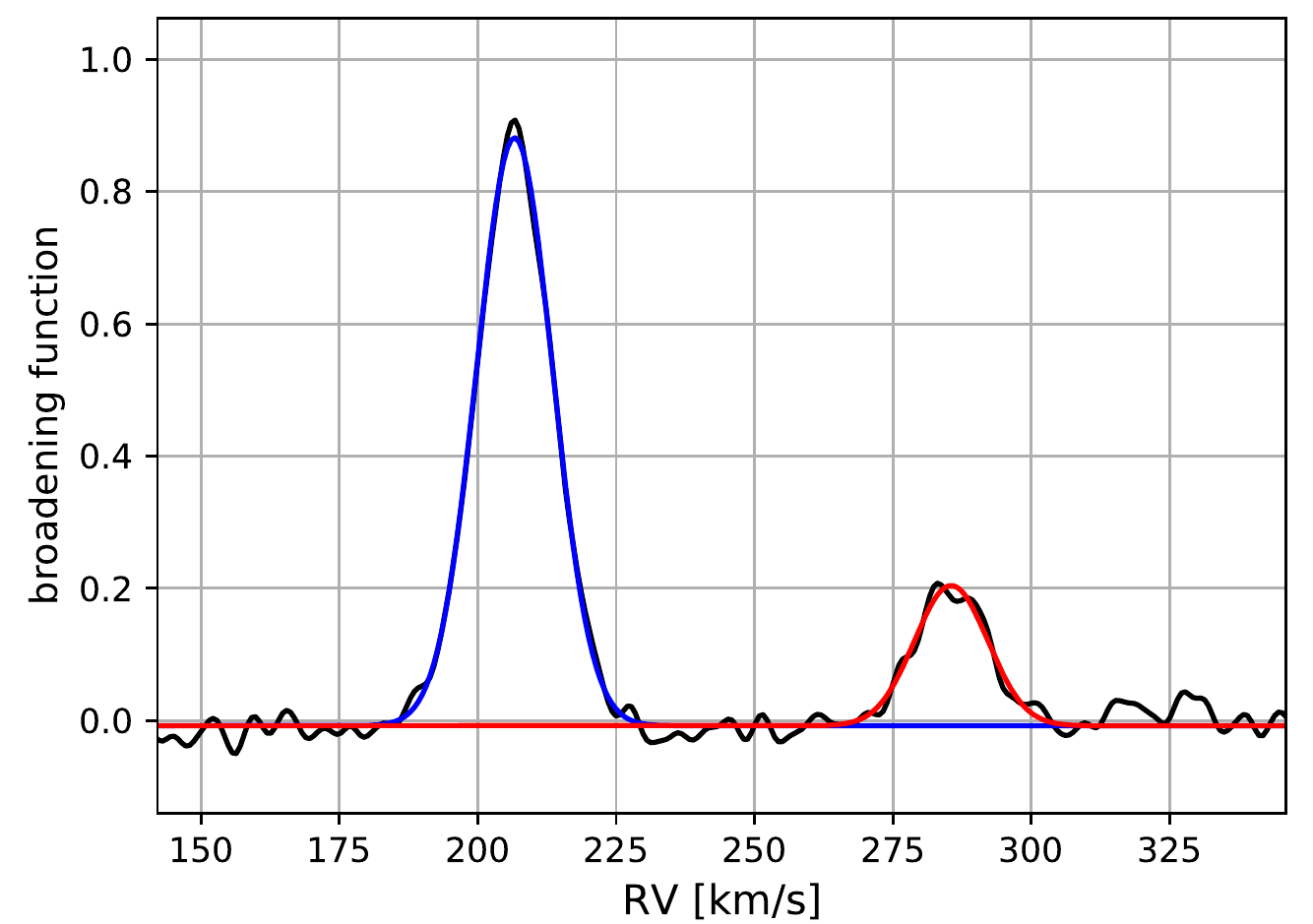}
    \end{center}
\caption{Broadening Function profiles of the Cepheid (blue) and the companion (red) for one of the UVES spectra. The signal from the Cepheid is $\sim$4 times stronger indicating the companion is significantly fainter (see text). }
\label{fig:bf}
\end{figure}

The BF profiles obtained for one of the UVES spectra are shown in Fig.~\ref{fig:bf}. The ratio of the integrated profiles is about 4. For stars of the same spectral type this would mean that the companion is roughly four times fainter over the wavelength range used for the BF measurement. The ratio of the integrated BF profiles is $\sim$5 for a wavelength range of 4100-5300 \AA, while the ratio is $\sim$3 for a wavelength range of 5300-6600 \AA, indicating that the companion is redder than the Cepheid.

\subsection{Orbital solution}
\label{sec:orbsol}

\begin{figure*}
    \begin{center}
        \includegraphics[width=0.57\textwidth]{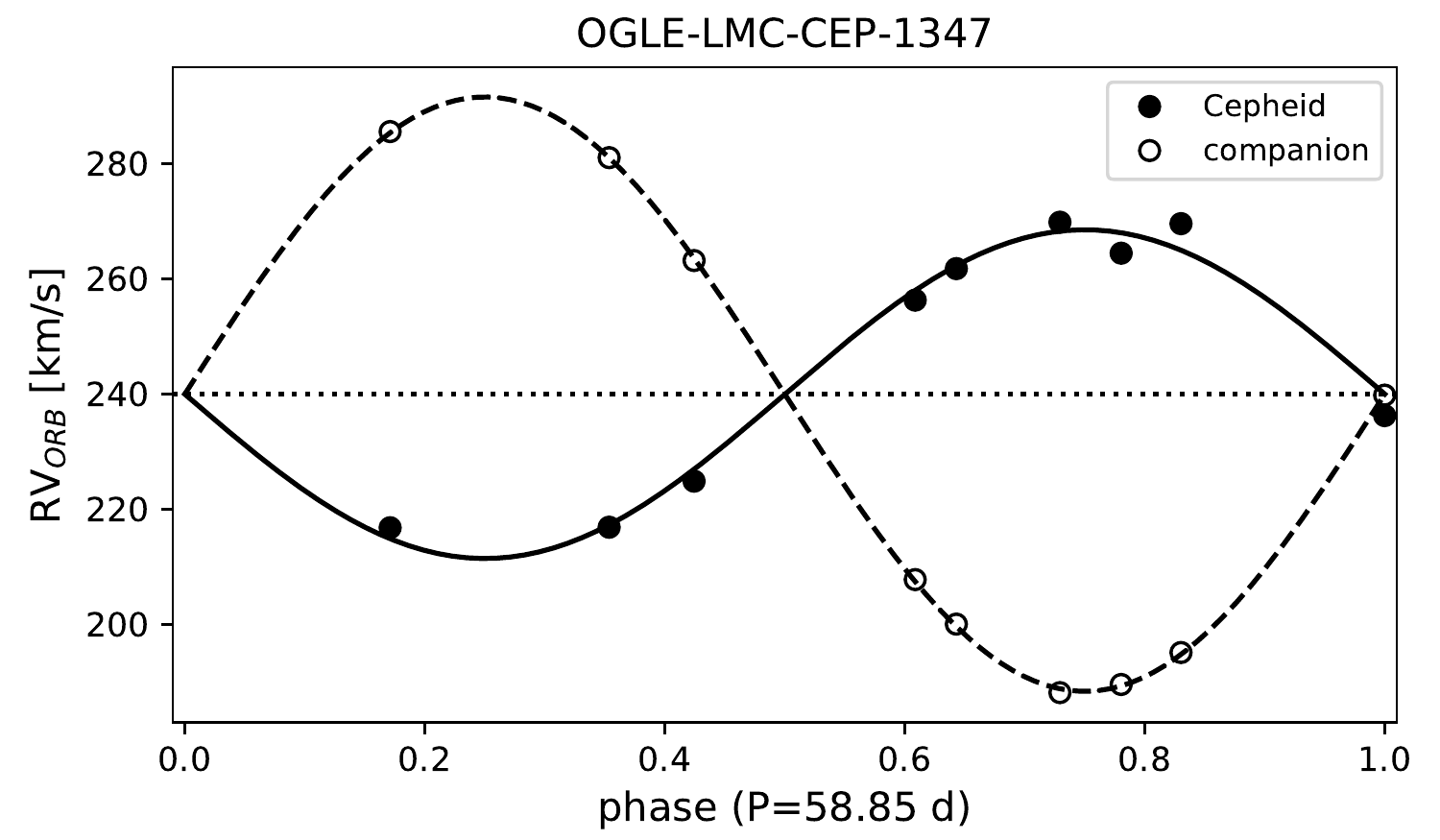}
        \includegraphics[width=0.42\textwidth]{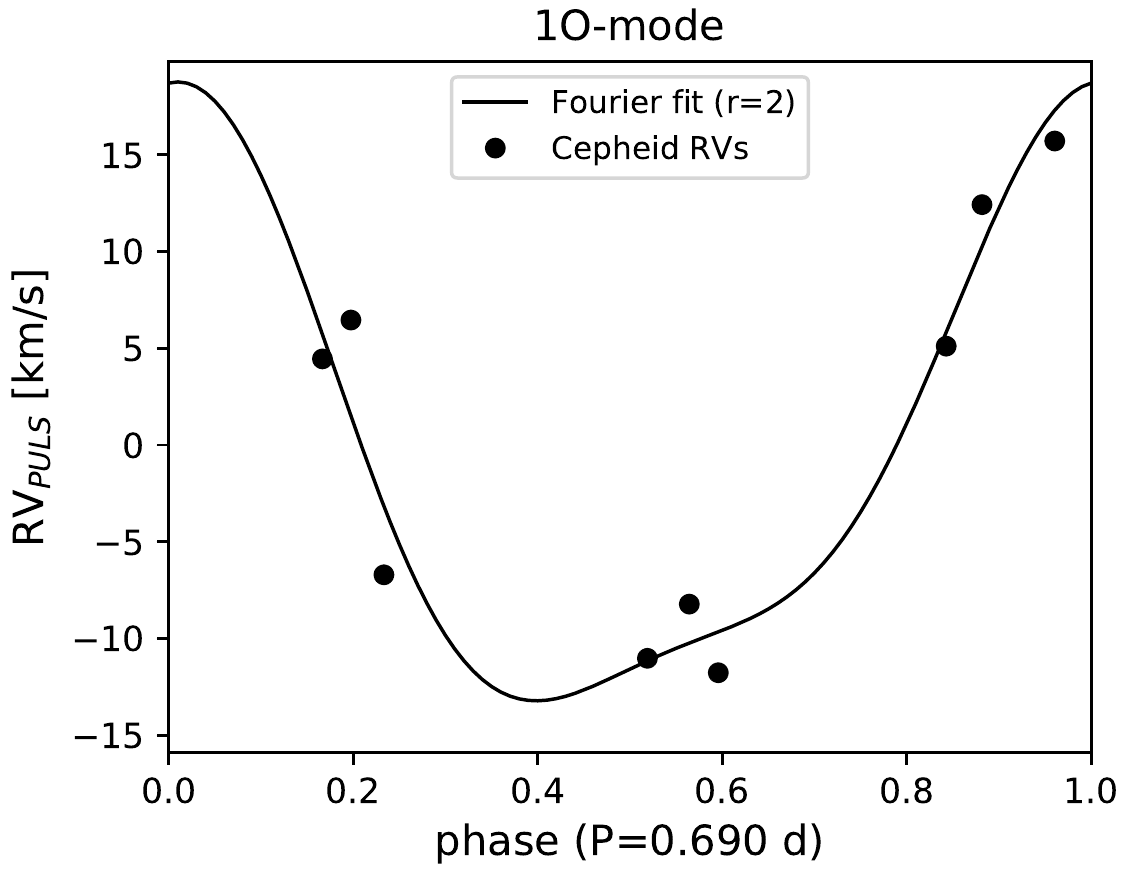}
    \end{center}
\caption{(left) Orbital radial velocity curves with a preliminary orbital solution. The $1O$-mode variation for the Cepheid was subtracted. The orbit is circular and different semi-amplitudes indicate the companion is about 2 times less massive. (right) Pulsational RV curve for $1O$ mode with orbital variation removed. High scatter of the Cepheid RVs around the fits comes from the unaccounted $2O$ pulsation. Measurement errors are smaller than the plotting points.}
\label{fig:orbsol}
\end{figure*}

To obtain the orbital solution we again used the RaveSpan code, adjusting the following orbital parameters: the orbital period ($P$), reference time ($T_0$), systemic velocity ($\gamma$), semi-amplitudes ($K_1$ and $K_2$), eccentricity ($e$) and the argument of periastron ($\omega$). For the current set of data the analysis yielded a circular orbit ($e$=0).
During the fitting the main pulsation variability with the period $0.690$ day was described by a 2nd-order Fourier series. The solution is presented in Table~\ref{tab:orb} and shown in Fig.~\ref{fig:orbsol}. In the table $a$ means semi-major axis, $i$ -- orbital inclination, $m_i$ -- mass of the i-th component and $q$ -- the mass ratio.
Because of the low number of observational points we did not take the $2O$ mode into account but its amplitude seems to be significantly lower than for the $1O$ mode with an rms value of 2.6 km/s for residual RVs compared to 9.8 km/s if neither mode is included in the fit. 

\begin{deluxetable}{lr@{ $\pm$ }lc}
\tablecaption{Preliminary orbital solution for OGLE-LMC-CEP-1347  \label{tab:orb}}
\tablewidth{0pt}
\tablehead{
\colhead{Parameter} & \multicolumn{2}{c}{Value} & \colhead{Unit}
}
\startdata
$P$             &   58.85    &  0.08   &  days   \\
$T_0$ (HJD)     & 2459050.0  &  0.7    &  days   \\
$a \sin i$      &   93.2     &  1.0    &  R$_\odot$ \\ 
$m_1 \sin^3 i$  &    2.02    &  0.04   &  M$_\odot$ \\ 
$m_2 \sin^3 i$  &    1.12    &  0.06   &  M$_\odot$ \\ 
$q=m_2/m_1$     &    0.553   &  0.016  &  -      \\ 
$e$             & \multicolumn2c{0.0}  &  -      \\
$\gamma$        & 239.97     &  0.13   &  km/s   \\ 
$K_1$           &   28.5     &  0.8    &  km/s \\   
$K_2$           &   51.56    &  0.17   &  km/s \\   
rms$_1$         & \multicolumn{2}{c}{2.61} &  km/s \\
rms$_2$         & \multicolumn{2}{c}{0.36} &  km/s 
\enddata
 \tablecomments{Only the statistical error is included. Second-overtone pulsation is not taken into account (note the high rms$_1$ value).}
\end{deluxetable}

The most striking feature of this system is the unexpectedly short orbital period, much shorter than for any other known binary Cepheid. A total of 1.5 orbital cycles are covered with a uniform distribution, so this detection is firm. The high scatter for the Cepheid RVs has a negligible effect here as the orbital period is constrained mainly by the precise RVs of the companion.
The orbital solution we obtained suggests a circular orbit and a high mass ratio of the components with the Cepheid being almost two times more massive than its companion. The minimum mass ($m \sin^3 i$) of the Cepheid is $\sim$2 M$_\odot$. Although the solution is still preliminary and the eccentricity and mass ratio may change slightly, these measurements already indicate some interesting aspects of the system.

\section{Discussion}
\label{sec:discussion}

Population synthesis done by \citet{Neilson2015_cep_binsys_porb} predicts that binary systems with Cepheids, with components having passed through the red giant phase, cannot have periods shorter than about 200 days. The reason is that binaries with shorter periods would have interacted. This is consistent with observations, as all binary Cepheids confirmed either spectroscopically or astrometrically, have periods of about 1 year or longer \citep{Evans_2013_orbperiods, allcep_pilecki_2018}. Although some candidates have apparent orbital periods shorter than 300 days \citep{Soszynski_2010_SMC_Cepheids}, the shortest period measured for a confirmed binary Cepheid is 310 days. That is why the very short period of only 59 days measured for LMC-CEP-1347 is so extraordinary. The most plausible explanation is that the Cepheid is on the first crossing through the instability strip and has not yet evolved onto the red giant branch. This scenario is supported by the observed rapid period increase $\log\dot{P_{1O}} = -1.00 \pm 0.03$ s/yr \citep{cepgiant1_2021} of the Cepheid. For the fundamentalized period ($P_F=0.966$ d) we obtain $\log\dot{P_F} = -0.85$ s/yr which matches theoretical predictions, e.g. $\log\dot{P_F}$ between -0.7 and -0.95 s/yr in \citet{Turner_2006_perchange_crossings}.
Moreover, at the LMC metallicity, modern evolutionary models usually do not predict the existence of Cepheids with such short pulsation periods after the red giant branch \citep{Anderson_2016_Rotation,DeSomma_2021_MNRAS_per-age}.

\begin{figure}
    \begin{center}
        \vspace{0.5cm}
        \includegraphics[width=1.0\linewidth]{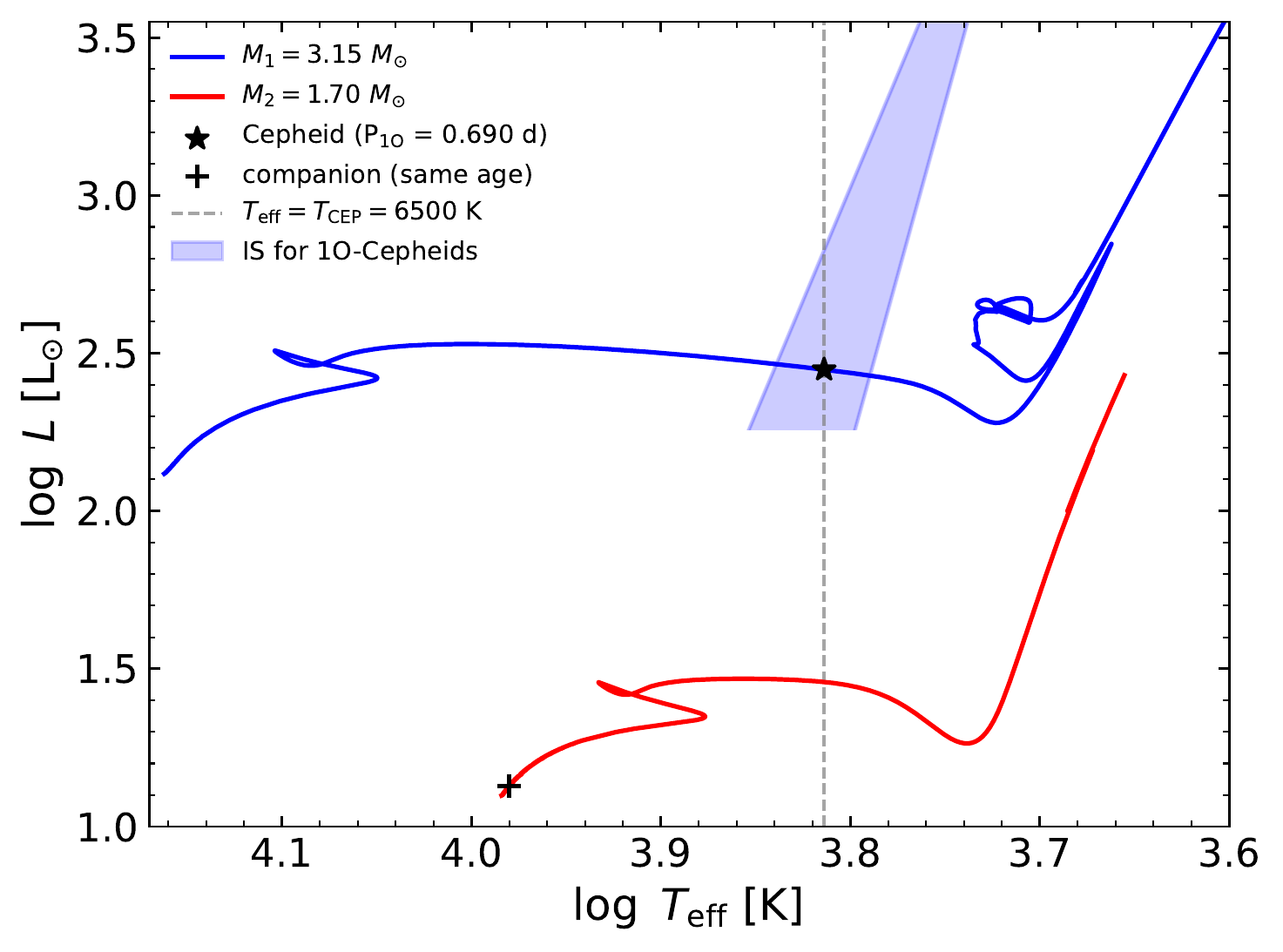}
    \end{center}
\caption{Example evolutionary tracks for stars with masses expected for components of OGLE-LMC-CEP-1347 assuming single star evolution. The position of the Cepheid with a $1O$-period of 0.690 day and the corresponding instability strip are shown. The hypothetical position of the companion with the same age as the Cepheid clearly contradicts the observational data (it should be cooler than the Cepheid). }
\label{fig:evol}
\end{figure} 

We calculated evolutionary models using the MESA code \citep{Paxton_2019ApJS_MESA} for a set of masses and metallicities ([Fe/H]=-0.54 to -0.27) consistent with that of the LMC \citep{Romaniello_2022_metal_LMC_Ceps}, and found the expected mass of a first-crossing Cepheid with a $1O$-period equal to that of LMC-CEP-1347 to be within 2.9--3.4 M$_\odot$. If observationally confirmed, this would be the lowest-mass Cepheid known. We also obtained expected period change rates $\log\dot{P_{1O}}$ between -0.74 and -1.00 s/yr.
Using the information from the orbital solution, the mass of the companion would be $\sim$1.7~M$_\odot$ and the orbital inclination around 60$^\circ$. Assuming the Cepheid is on a higher crossing, we would obtain an even lower mass at which, however, our models do not predict blue loops that enter the instability strip. The evolutionary tracks were calculated taking into account diffusion, semi-convection, thermohaline mixing, and mass loss. We used a mixing length of 1.88 and overshooting mixing efficiency of 0.019, both solar-calibrated. The boundaries of the instability strip were obtained applying the MESA Radial Stellar Pulsations package \citep{sm08} to the evolutionary models.

The preliminary mass ratio of the components is far from unity, while the stars are apparently in a similar evolutionary stage as indicated by their spectral features and comparable brightness. Indeed, subtracting the typical Cepheid brightness obtained from the period-luminosity relations \citep{Soszynski_2008_LMC_Cepheids} from the observed brightness in $V$ and $I$-bands, we find that the companion is fainter ($\Delta V$=0.64 mag) and redder ($\Delta (V-I)$=0.17 mag) than the Cepheid, which supports the conclusion from the analysis of BF profiles in Section~\ref{sec:spectro}.
This means that assuming single star evolution the companion would have to be at a similar or even later evolutionary stage than the Cepheid, while having the same age as the Cepheid it would be still on the main sequence as seen in Fig.~\ref{fig:evol}. It would be also much bluer (hotter) than the Cepheid.
All the observational facts (including the mass ratio and the color difference) can be explained if we assume that LMC-CEP-1347 has been a triple system, with the components of the inner binary having merged creating the current Cepheid with a mass about twice that of the companion. A similar scenario was proposed for the binary Cepheid OGLE-LMC-CEP-1812 \citep{Neilson2015_cep1812_merger} with the mass ratio of about 0.7 and components also at a similar evolutionary stage. Another possibility is mass transfer from the current secondary as in \citet{t2cep211_pilecki2018}. However in that case we would expect either the current rejuvenated primary (more massive component) to still be on the main sequence or the evolutionarily advanced secondary (less massive component), stripped of the envelope, to be hardly detectable in the spectra.

The short orbital period and low component separation means the Cepheid may in the future again interact, this time with its current companion, either by merging or mass transfer. This is a hint that for some Cepheids the first crossing may also be the last one. As there are many B-type binary systems with similarly short orbital periods \citep{Duchene_2013_multip_review,Moe_2017_multip_review}, the number of first-crossing Cepheids may be higher than typically assumed from a comparison of evolutionary time spent within the instability strip at different crossings \citep[e.g.][]{Ripepi_2022_MNRAS_VMC_LMC_CEPs}.

Apart from being an example of the complexity of interactions in multiple systems of intermediate and high-mass stars, LMC-CEP-1347 has some other impressive features. It does not only belong to the binary system with about a five times shorter orbital period than previously known for binary Cepheids, but is also the shortest-period Cepheid found in a binary system so far. Although other double-mode Cepheids in binary systems have been identified \citep[e.g., Y Car; ][]{Evans_1992_double_mode_cep_YCar}, LMC-CEP-1347 is also the first one to be found in an SB2 system. Double-mode Cepheids provide a very robust, purely photometric method to constrain Cepheid masses and luminosities through their period ratios \citep{Petersen_1973_mass_period_ratio, Bono_1996_rrlyr_mass_period_ratio}, i.e., independently of the distance and reddening. Unfortunately, this method has never been calibrated empirically because of a lack of dynamical mass measurements for double-mode Cepheids. LMC-CEP-1347 may thus be the first to be used for that purpose once more data is acquired.


\acknowledgments
The research leading to these results has received funding from the Polish National Science Center grant SONATA BIS 2020/38/E/ST9/00486. We also acknowledge support from the European Research Council (ERC) under the European Union's Horizon 2020 research and innovation program (grant agreement No. 951549). W.G. gratefully acknowledges support from the ANID BASAL project ACE210002.
This work is based on observations collected at the European Southern Observatory under ESO program 108.22BS.001. We also thank Carnegie, and the CNTAC for the generous allocation of observing time for this project. We would like to thank the support staff at the Las Campanas Observatory for their help in remote observations.

This research has made use of NASA's Astrophysics Data System Service.

\vspace{5mm}
\facilities{VLT:Kueyen (UVES), Magellan:Clay (MIKE)}

\software{
\texttt{RaveSpan} \citep[][\url{https://users.camk.edu.pl/pilecki/ravespan/}]{t2cep098apj2017}
\texttt{MESA} \citep[][\url{https://docs.mesastar.org/}]{Paxton_2019ApJS_MESA}
}



\bibliography{lmc-cep-1347}{}
\bibliographystyle{aasjournal}



\end{document}